\documentstyle[12pt]{article}
\begin{document}
\begin{flushright}
{IHEP 95-81}\\
hep-ph/9507224\\
\end{flushright}
\vspace*{4mm}
\begin{center}
{\large \bf Modified renormalon}\\
\vspace*{4mm}
V.V.Kiselev\\
Institute for High Energy Physics,\\
Protvino, Moscow Region, 142284, Russia.\\
E-mail: kiselev@mx.ihep.su\\
Fax: +7-(095)-230 23 37.
\end{center}
\begin{abstract}
In the framework of renormalon consideration in the approximation of a large
number of quark flavours, a role of the anomalous dependence of gluon
propagator on the scale $\mu$ is shown.
\end{abstract}

In the QCD perturbation theory, an account for the next-to-leading order
term over $\alpha_s$ is connected with the uncertainty, caused by a choice
of the energy scale, determining the $\alpha_s(\mu^2)$ value, since the
transition to the $\bar \mu$ scale leads to the substitution
$$
\alpha_s(\mu^2) \approx \alpha_s(\bar \mu^2)\biggl(1 - \alpha_s(\bar \mu^2)
\frac{\beta_0}{4\pi} \ln\frac{\mu^2}{\bar \mu^2}\biggr)\; ,
$$
where $\beta_0 =(11 N - 2n_f)/3$, $N$ is the number of colours, $n_f$ is
the number of quark flavours. So, the physical quantity, represented in the
form
$$
r = r_0 \alpha_s(\mu^2)\{1+ r_1 \alpha_s + O(\alpha_s^2)\}
$$
with the given order of accuracy, will get the form
\begin{eqnarray}
r & = & r_0 \alpha_s(\bar \mu^2)\bigg\{1+ \biggl(r_1 - \frac{\beta_0}{4\pi}
\ln\frac{\mu^2}{\bar \mu^2}\biggr) \alpha_s + O(\alpha_s^2)\bigg\} \nonumber\\
&=& r_0 \alpha_s(\bar \mu^2)\{1+\bar r_1 \alpha_s + O(\alpha_s^2)\}\;.\nonumber
\end{eqnarray}
Thus, the value of $r_1$ coefficient for the $\alpha_s$ correction depends on
the scale of $\alpha_s(\mu^2)$ determination. In ref.\cite{blm}, one
offered to fix the $\mu$ scale so that the $r_1$ coefficient does not
contain terms, proportional to $\beta_0$. The latter procedure can be
understood as one, leading to $r_1$ to be independent of the number of
quark flavours $n_f \sim \beta_0$. As was shown in refs.\cite{2,3}, such
choice of scale has the strict sense in the framework of the $1/n_f$ expansion,
where the $\alpha_s$ correction to the gluon propagator is determined by
the fermion loop contribution into the vacuum polarization. So, the expression
for such contribution depends on the regularization scheme, and in the
next-to-leading order one has
\begin{equation}
\alpha_s(\mu^2) D(k^2, \mu) = \frac{\alpha_s(\mu^2)}{k^2}\;
\bigg\{1+ \frac{\alpha_s n_f}{6\pi}
\biggl(\ln\frac{k^2}{\mu^2} + C\biggr)\bigg\}\;,
\label{1}
\end{equation}
where $D(k^2,\mu)$ determines the transversal part of the gluon propagator
$D^{ab}_{\nu\lambda}(k^2,\mu) = \delta^{ab} D(k^2,\mu) (-g_{\nu\lambda}+
k_\nu k_\lambda/k^2)$. The constant value $C$ is defined by the renormalization
scheme, so $C_{\overline{MS}} = -5/3$ and $C_V =0$ in the so-called $V$-scheme
\cite{blm}.

The summing of $(n_f\alpha_s)^n$ contributions into the gluon propagator
leads to the expression
\begin{equation}
\alpha_s(\mu^2) D(k^2, \mu) = \frac{\alpha_s(e^C k^2)}{k^2}\;,
\label{g1}
\end{equation}
i.e., in fact, it results in the account for the "running" $\alpha_s$ value in
respect to the gluon virtuality. Note, the $1/n_f$ consideration is exact
for the abelian theory, where $\alpha_s D$ is the renormalization group
invariant. Moreover, the $e^{-C}\Lambda^2_{QCD}$ value does not depend on the
renormalization scheme, and therefore it results in the scheme-independence
for the expression
$$
\alpha_s(e^C k^2) = \frac{4\pi}{\beta_0\ln(e^C k^2/\Lambda^2_{QCD})}\;.
$$

As one can see in the performed consideration, the transition to the
nonabelian theory is done by the $2n_f/3 \to -\beta_0$ substitution, that is
called as the procedure of "naive nonabelization", giving the correct
"running" of the coupling constant.

Further, consider the physical quantity $r$, for which the first order
$\alpha_s$-contribution is calculated as an integral over the gluon
virtuality with the weight $F(k^2)$
$$
r = \int \frac{dk^2}{k^2}\; \alpha_s(\mu^2) F(k^2)\;.
$$
Then the account for the $(n_f\alpha_s)^n$ corrections to the gluon propagator
leads to the substitution of "running" value $\alpha_s(e^C k^2)$ for
$\alpha_s(\mu^2)$
\begin{equation}
\alpha_s(\mu^2) \to \alpha_s(\mu^2)\sum_{n=0}^{\infty}
\bigg\{-\frac{\alpha_s(\mu^2)\beta_0}{4\pi}\biggl(\ln\frac{k^2}{\mu^2}+C\biggr)
\bigg\}^n\;,
\label{2}
\end{equation}
so that in the offered procedure of the scale fixing \cite{blm}, the
introduction of the first order $n_f\alpha_s$-correction results in the
substitution
$\mu^2\to \bar \mu^2 = \mu^2 \exp\{C+\langle\ln k^2/\mu^2\rangle\}$, where
the average value is determined by the integral with the $F$ weight.
However, in some cases \cite{2,3} the integration of the $n$-th order term
in expansion (\ref{2}) leads to the $n!$ factorial growth of the coefficients
for the expansion over $\alpha_s^n(\mu^2)$, so that this rising can be
determined by the region of low virtualities as well as large ones of gluon.
Then one says about the infrared and ultraviolet renormalons, respectively
\cite{2,3,4}. As was shown \cite{2,3}, the renormalon leads to power-like
uncertainties of the $r$ evaluation, so
\begin{equation}
\Delta r = \biggl(\frac{\Lambda_{QCD}}{\mu}\biggr)^k a_k\;,
\label{3}
\end{equation}
where $k$ is positive for the infrared renormalon and it is negative for the
ultraviolet one. For the two-point correlator of heavy quark vector currents,
for instance, one has $k=4$ and the corresponding uncertainty can be, in fact,
eliminated in the procedure of definition for the nonperturbative gluon
condensate, having the same power over the infrared parameter \cite{3}.
However,  for the renormalized mass of heavy quark the infrared renormalon
with $k=1$ does not correspond to some definite quark-gluon condensate, and
the value $\Delta m(\mu) \sim \Lambda_{QCD}$ can not be adopted into a
definition of a physical condensate.

In the present paper we modify the renormalon due to the account for the
anomalous dependence of gluon propagator on the $\mu$ scale, and the
modification turns out, basically, to be essential for numerical estimates,
related with the determination of physical scale, fixing $\alpha_s$.

In the covariant gauge, the gluon propagator has the form
$$
D^{ab}_{\nu\lambda}(k^2,\mu) = \frac{\delta^{ab}}{k^2\omega(\mu^2,k^2)}
\biggl(-g_{\nu\lambda}+(1-a_l(\mu^2)\omega(\mu^2,k^2))
\frac{k_\nu k_\lambda}{k^2}\biggr)
\;,
$$
where $a_l\omega$ is the renormalization group invariant. The solution of
one-loop equation of the renormalization group for $a_l(\mu^2)$ in the
$\overline{MS}$-scheme allows one to represent the gluon propagator in the form
\cite{kis}
\begin{equation}
D^{ab}_{\nu\lambda}(k^2,\mu) = \frac{\delta^{ab}}{k^2}
\frac{1}{1-(1-\omega(\mu_0^2))\biggl(\frac{\alpha_s(\mu^2)}
{\alpha_s(\mu_0^2)}\biggr)^{n_a}}
\biggl(-g_{\nu\lambda}+(1-b)
\frac{k_\nu k_\lambda}{k^2}\biggr)\;,
\label{4}
\end{equation}
where $n_a=(13 N-4 n_f)/(6\beta_0)$ and $b$ is the arbitrary gauge
parameter\footnote{In ref.\cite{kis} the solution with the fixed $b=3$ value is
considered and $\omega$ is expressed through $b/a_l$.}.
In eq.(\ref{4}) we take into account the anomalous dependence for the gluon
propagator on the scale, $\omega=\omega(\mu^2)$, only.

The choice of fixed point $\omega \equiv 1$ corresponds to the standard
renormalon. The consideration of the gluon propagator at $\omega\neq 1$
leads to the modified renormalon.

In the leading order over $1/n_f$ one has $n_a=1$, and the cases with
1) $\omega(\mu_0^2)>1$, 2) $\omega(\mu_0^2)=1$ and 3) $\omega(\mu_0^2)<1$
at some large $\mu_0$ values can be formally come to the different choices
of the renormalization group invariant $\mu_g$, such that $\omega(\mu_g^2)=0$
and 1) $\mu_g<\Lambda_{QCD}$, 2)  $\mu_g=\Lambda_{QCD}$ and 3)
$\mu_g>\Lambda_{QCD}$, respectively. The $\alpha_s(\mu_g^2)$ value
is considered as the formal one-loop expression
$$
\alpha_s(\mu^2) = \frac{4\pi}{\beta_0\ln\frac{\mu^2}{\Lambda^2_{QCD}}}\;.
$$
Thus, the $1/n_f$ consideration gives
\begin{equation}
\frac{\alpha_s(\mu^2)}{\omega(\mu^2)} = \frac{\alpha_s(\mu^2)
\alpha_s(\mu_g^2)}{\alpha_s(\mu_g^2) - \alpha_s(\mu^2)} =
\alpha_s(e^d\mu^2)\;,
\label{5}
\end{equation}
where $d$ is the scheme-independent invariant of the one-loop renormalization
group, and it is equal to
$$
d=\ln\frac{\Lambda^2_{QCD}}{\mu_g^2}=
-\frac{4\pi}{\beta_0 \alpha_s(\mu_g^2)}\;.
$$
Further, one can repeat the calculation of $(n_f\alpha_s)^n$ contributions
into the gluon propagator with the substitution of value (\ref{5}) for
$\alpha_s(\mu^2)$, so that this replacement corresponds to the account for
the anomalous dependence of the gluon propagator versus the $\mu$ scale.
Then the scale fixing due to the next-to-leading order $\alpha_s$ correction
results in the expression
$$
\bar \mu^2 = \mu^2 \exp\{C+\langle\ln k^2/\mu^2\rangle+d\}\;,
$$
and the summing of the corresponding contributions modifies eq.(\ref{g1})
\begin{equation}
\alpha_s(\mu^2) D(k^2, \mu) = \frac{\alpha_s(e^{C+d} k^2)}{k^2}=
\frac{\alpha_s(e^C k^2)}{\omega(e^C k^2)k^2}\;.
\label{g2}
\end{equation}

Next, one can define the $\overline{V}$-scheme, introducing
$C_{\overline{V}}=-d$. Then
$$
\Lambda^{\overline{V}}_{QCD} = e^{5/6-d/2}\Lambda^{\overline{MS}}_{QCD}\;.
$$
In the $\overline{V}$-scheme the perturbative potential between the
heavy quark and antiquark in the colour-singlet state will have the form
\begin{equation}
V(q^2) = - \frac{4}{3}\; \frac{4\pi\alpha_s^{\overline{V}}(q^2)}{q^2}
\label{v}
\end{equation}
at $q^2 = {\bf k}^2$. Potential (\ref{v}) comes to the Richardson potential
\cite{rich}, when one uses $\alpha_s^{\overline{V}}(q^2+\Lambda^2_{QCD})$
instead of $\alpha_s^{\overline{V}}(q^2)$, and $\Lambda_{QCD}$ fixes the
linearly rising part of potential, confining quarks with the distance increase,
$$
\Delta V_{lin}({\bf x}) = \frac{8\pi\Lambda^2_{QCD}}{27} |{\bf x}|\;.
$$
The fitting of mass spectra for the charmonium and bottomonium in the
Richardson potential gives
$$
\Lambda^{\overline{V}}_{QCD} = 400\pm 15\; \mbox{MeV}\;.
$$
The one-loop expression with $\alpha_s^{\overline{MS}}(m_Z^2) = 0.117\pm 0.005$
\cite{pdg}
corresponds to $\Lambda^{\overline{MS}}_{QCD} = 85\pm 25\; \mbox{MeV}$,
that allows one  phenomenologically to determine the values
\begin{eqnarray}
e^{-d/2} & = & 2.1\pm 0.5\;,\nonumber \\
\mu_g^{\overline{MS}} & = & 180\pm 45\; \mbox{MeV}\;,\label{e} \\
\alpha_s(\mu_g^2) & = & 0.94\pm 0.21\;. \nonumber
\end{eqnarray}
However, an additional uncertainty into the estimates is involved by the
transition from the number of flavours $n_f(m_Z)=5$ to another value
$n_f(1\; \mbox{GeV})=3$. Obtaining eq.(\ref{e}), one has supposed, that
$\Lambda_{QCD}$ does not depend on the number of flavours, so that at the
scale of "switching on (off)" the additional flavour of quarks $\mu=m_{n_f+1}$,
the QCD coupling has a discontinuity, related with the step-like change
$\beta_0(n_f)\to \beta_0(n_f+1)=\beta_0(n_f)-2/3$. One can avoid such
discontinuities, if one supposes that $\Lambda_{QCD}$ depends on the
flavour number, so that $\alpha_s(\mu^2=m^2_{n_f+1},\Lambda^{(n_f)}_{QCD},
n_f)=\alpha_s(\mu^2=m^2_{n_f+1},\Lambda^{(n_f+1)}_{QCD},n_f+1)$.
Then in the one-loop approximation one finds
\begin{equation}
\Lambda^{(n_f)}_{QCD} = \Lambda^{(n_f+1)}_{QCD}\biggl(\frac{m_{n_f+1}}
{\Lambda^{(n_f+1)}_{QCD}}\biggr)^\frac{2}{3\beta_0(n_f)}\;.
\label{l}
\end{equation}
Setting $\Lambda^{(5)}_{QCD}=85\pm 25$ MeV, one gets
\begin{equation}
e^{-d/2} = 1.24\pm 0.32\;. \label{e2}
\end{equation}
Hence, the $\alpha_s$ rescaling, accounting for eq.(\ref{l}), results in
the $d$ value, that agrees with the standard renormalon ($d\equiv 0$)
within the current accuracy.

Next, one can use the nonabelian "running" of $\omega$ with $n_a\neq 1$
and the same scale as in the "running" of $\alpha_s$. In the latter
procedure one has two scales $\Lambda_{QCD}$ and $\mu_g$, so that two
"measurements" of $\alpha_s/\omega$ quantity, entering physical values,
are needed. If one fixes $\mu_g^V$ from the linear part of the heavy quark
potential and uses the $Z$ pole data for the $\Lambda_{QCD}$ evaluation,
then the obtained values show the standard renormalon validity within errors.

Thus, we have found that in the $1/n_f$ consideration the account for the
anomalous dependence of the gluon propagator on the $\mu$ scale results in
the modification of renormalon because of the presence of
$\mu_g$ scale in addition to $\Lambda_{QCD}$, and the difference from
the standard renormalon turns out to be essential in numerical estimates.
We have also determined $\mu_g^{\overline{MS}}$ in a phenomenological manner.

This work is partially supported by the International Science Foundation
grants NJQ000, NJQ300 and by the program "Russian State Stipendia for young
scientists".

\vspace*{0.4cm}

\hfill {\it Received June 9, 1995}
\end{document}